\documentclass{article}
\usepackage{spconf,amsmath,amsfonts,graphicx,bm}
\usepackage{caption,subcaption,pgfplots}
\usepackage{hyperref}
\usepackage{multirow}

\title{Lightweight Speech Enhancement in Unseen Noisy and Reverberant Conditions using KISS-GEV Beamforming}
\name{Thomas Bernard, Fran\c{c}ois Grondin\thanks{This work was supported by the Natural Sciences and Engineering Research Council of Canada (NSERC) and Mitacs. This work was done in collaboration with Stratégie Agilean inc.}}
\address{Universit\'e de Sherbrooke, Sherbrooke (Québec), Canada} 
\begin{document}
%
\maketitle
\begin{abstract}
This paper introduces a new method referred to as KISS-GEV (for Keep It Super Simple Generalized eigenvalue) beamforming. While GEV beamforming usually relies on deep neural network for estimating target and noise time-frequency masks, this method uses a signal processing approach based on the direction of arrival (DoA) of the target. This considerably reduces the amount of computations involved at test time, and works for speech enhancement in unseen conditions as there is no need to train a neural network with noisy speech. The proposed method can also be used to separate speech from a mixture, provided the speech sources come from different directions. Results also show that the proposed method uses the same minimal DoA assumption as Delay-and-Sum beamforming, yet outperforms this traditional approach.
\end{abstract}
\section{Introduction}
\label{sec:intro}

Distant speech processing is a challenging task as the target speech is usually corrupted by reverberation and interfering noise sources from the environment \cite{tang2018study}. 
Training an automatic speech recognition (ASR) system in the same environmental conditions is desirable, yet remains a challenging task when the additive noise and reverberation to be observed at test time is unknown \cite{shimada_unsupervised_2019}.
Enhancing speech with a microphone array prior to the ASR task stands as a simple solution to cope with the training and testing conditions mismatch \cite{chen_building_2018}.

Delay-and-Sum (DS) and Minimum Variance Distortionless Response (MVDR) beamformers use the direction of arrival (DoA) of the target sound source to enhance speech \cite{chen2018multi, habets2009new, erdogan2016improved, xiao2017time}.
However, they rely on the anechoic assumption, i.e. free field propagation of sound, which is an inaccurate approximation for reverberant environments.
Alternatively, generalized eigenvalue (GEV) beamforming can enhance speech using only estimations of the speech and noise spatial covariance matrices (SCMs), based on an estimated time-frequency mask to distinguish speech from background noise.
A neural network usually estimates this time-frequency mask, which implies the background interference type has to be known in advance at training time \cite{heymann2015blstm, heymann2016neural, heymann2017beamnet, grondin2020gev}. While this can yield satisfactory performance when trained and evaluated on large synthetic datasets \cite{grondin2020bird}, this approach remains challenging in real environments with unknown interference.

The proposed method, referred to as KISS-GEV (for Keep It Super Simple GEV), assumes the DoA of the target signal is obtained using a sound localization method \cite{grondin2019svd} or audio-visual fusion \cite{grondin2020audio}.
The idea behind KISS-GEV is to provide a mask estimation method that is computationally much lighter than state of the art machine learning-based approaches, while providing adequate enhancement in real-world scenarios with unseen noise, making it viable to run on low-cost embedded hardware.
Results demonstrate that this method works without training a neural network to predict a time-frequency mask to differentiate speech from noise, which makes it ideal for enhancement in unseen conditions.
The proposed method can also be used to separate speech from a mixture, provided the speech sources come from different directions.
Results also show that the proposed method uses the same minimal DoA assumption as DS beamforming, yet outperforms this traditional approach.

\section{Proposed approach}

\begin{figure*}[!ht]
    \centering
    \includegraphics[width=\linewidth]{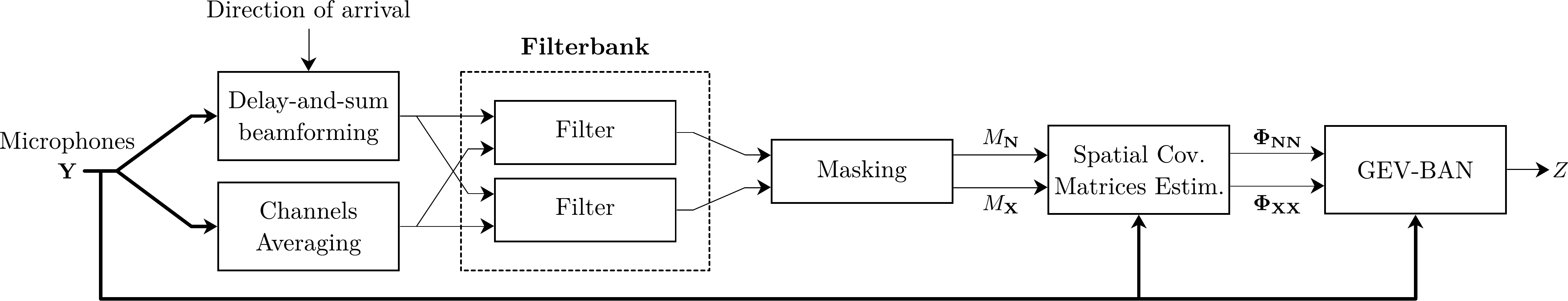}
    \caption{Overview of the KISS-GEV processing pipeline}
    \label{fig:overview}
\end{figure*}

KISS-GEV compares the power of the signal obtained with DS beamforming pointing in the direction of the target against the average power from all channels.
A two-filter bank then computes the power in low and high frequency regions, and generates a coarse binary time-frequency mask representing the target signal.
These two broadband filters are suitable for reverberant environments as they can cope with early reflections, which usually significantly impact the spectral shape of the DS beamformed signal.
This idea is similar to the Generalized Cross-Correlation with Phase Transform (GCC-PHAT) approach, where analyzing the normalized broadband spectrum makes time difference of arrival (TDoA) estimation more robust to early reflections \cite{brandstein1997robust}.
With $\mathbf{X}$ representing the target and $\mathbf{N}$, the interference, the masks obtained with the filterbank ($M_{\mathbf{X}}$ and $M_{\mathbf{N}}$) then provide an estimation of the target and interference spatial covariance matrices, denoted as ${\Phi}_{\mathbf{XX}}$ and ${\Phi}_{\mathbf{NN}}$, respectively.
The two SCMs are finally used to perform GEV beamforming.
Figure \ref{fig:overview} illustrates this pipeline.

\begin{figure}[!h]
    \centering
    \includegraphics[width=\linewidth]{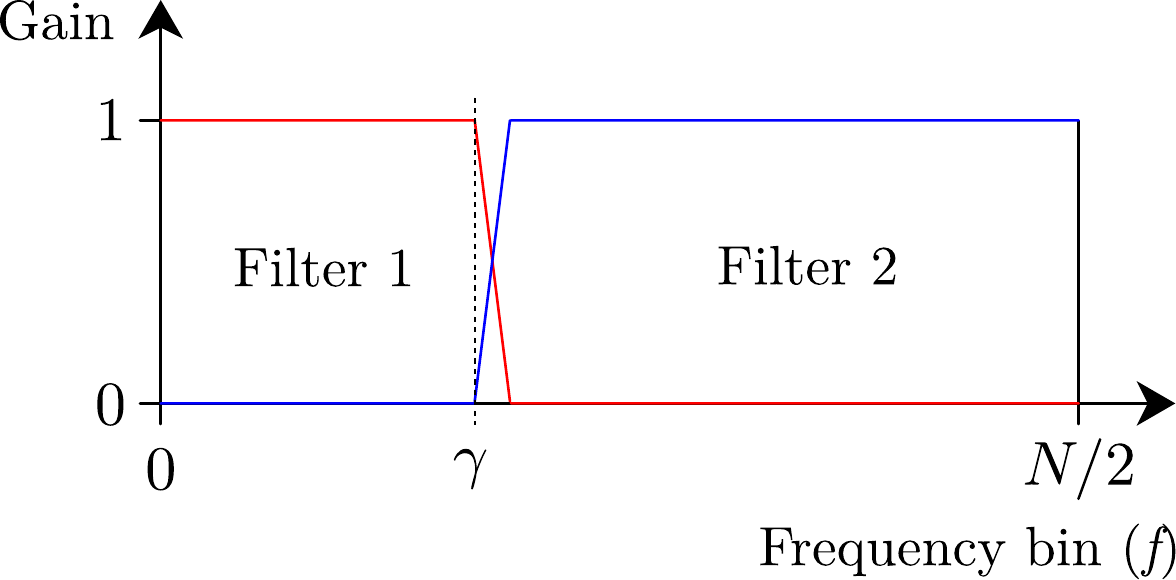}
    \caption{Two-filter filterbank, with first filter ($b=1$) in red, and second filter ($b=2$) in blue.}
    \label{fig:filterbank}
\end{figure}

The Short-Time Fourier Transform (STFT) is computed for the signal captured by each microphone $d \in \{1, 2, \dots, D\}$ (where $D$ denotes the number of microphones) to generate the time-frequency frames $Y_d(t,f) \in \mathbb{C}$, with frames of $N \in 2\mathbb{N}$ samples in the time-domain, where $t \in \mathbb{N}$ stands for the frame index and $f \in \{0, 1, \dots, N/2\}$ the frequency bin index.
The proposed approach then defines a binary filterbank made of $B \in \mathbb{N}$ filters, where $b \in \{1, 2, \dots, B\}$ is the filter index:

\begin{equation}
    H_b(f) = \begin{cases}
    1 & l_b \leq f \leq u_b \\
    0 & \mathrm{otherwise}
    \end{cases},
    \label{eqn:filterbank}
\end{equation}
where $l_b, u_b \in \{0, 1, \dots, N/2\}$ represent the lower and the upper bounds of the filter $b$, respectively.
In this paper, we assume there is no overlap between adjacent filters ($u_b < l_{b+1}$), and the spectrum is fully covered by the filters in the filterbank ($l_1 = 0$, $l_{b+1} = u_b + 1$ and $u_B = N/2$).
More specifically, we restrict the number of filters to $B=2$ (to intuitively capture either low-frequency voiced phonemes or high-frequency fricatives), which implies the only parameter of the filterbank is the separator position $\gamma = l_2 = u_1 + 1$, as shown in Figure \ref{fig:filterbank}.

The delay-and-sum beamformer is computed using the anechoic steering vector $W_d(f) \in \mathbb{C}$ towards the target:
\begin{equation}
    W_d(f) = \exp{\left(j2\pi f\tau_d/N\right)},
\end{equation}
where $\tau_d \in \mathbb{R}$ is the Time Delay of Arrival in samples, obtained from the known array geometry and target DoA.

The power of the beamformed signal corresponds to $|\sum_d W_d(f) Y_d(t,f)|^2$ for each frequency $f$, while the sum of the power of each channel is $\sum_d |Y_d(t,f)|^2$.
It is then possible to compute the total beamformed power, as well as the average power, for each filter $b$ in the filterbank, by multiplying by the expression $H_b(f)$ and summing across all frequencies.
The ratio $R_b(t)$ of the beamformed and average power for each filterbank is hence defined as:
\begin{equation}
    R_b(t) = \frac{\sum_{f}{H_b(f)|\sum_{d}{W_d(f)Y_d(t,f)}|^2}}{D\sum_{f}{H_b(f)\sum_{d}{|Y_d(t,f)|^2}}}.
\end{equation}
where the constant $D$ is introduced for normalization purpose.

In other words, when the frequency region covered by the filter $b$ is dominated by the target signal, we expect the ratio to get closer to a value of $1$.
Similarly, when there is little power from the target signal in the region covered by filter $b$, the ratio gets closer to a value of $0$.
The complete time-frequency ratio for frame $t$ and bin $f$ is then obtained by summing every $R_b$ with the filterbank:
\begin{equation}
    R(t,f) = \sum_{b}{R_b(t)H_b(f)}.
\end{equation}

Then, only both extremities of $R(t,f)$, with a width defined by parameter $\alpha$, are kept to define binary masks in order to capture only the most significant spectral features. Thresholds $T_{\nu}(f)$ are defined as follows:
\begin{equation}
    T_{\nu}(f) = \begin{cases}
    P_{100-\alpha}(R(f)) & \mathrm{if}\ \nu = \mathbf{X}\\
    P_{\alpha}(R(f)) & \mathrm{if}\ \nu = \mathbf{N}\\
    \end{cases},
\end{equation}
where $P_i$ is the $i$-th percentile of $R$ for bin $f$ across all $t$.

Using these thresholds, the binary time-frequency masks $M_{\mathbf{\nu}}(t,f)$ are defined as:
\begin{equation}
    M_{\nu}(t,f) = \begin{cases}
    1 & \mathrm{if}\ R(t,f) > T_\mathbf{X}(f), \nu = \mathbf{X}\\
    1 & \mathrm{if}\ R(t,f) < T_\mathbf{N}(f), \nu = \mathbf{N}\\
    0 & \mathrm{otherwise}
    \end{cases},
\end{equation}

With the masks defined, both SCMs can be estimated as follows:
\begin{equation}
    \bm{\Phi}_{\nu\nu}(f) = \sum_{t}{M_{\nu}(t,f)\mathbf{Y}(t,f)\mathbf{Y}(t,f)^{\mathrm{H}}},
\end{equation}
where $\mathbf{Y}(t,f) \in \mathbb{C}^{D \times 1}$ stands for a vector that concatenates the time-frequency frames $Y_d(t,f) \in \mathbb{C}$ of all microphones, and $(\dots)^H$ stands for the Hermitian operator.

The beamforming vector $\mathbf{F}_{\mathrm{GEV}}(f) \in \mathbb{C}^{D \times 1}$ is then obtained by performing the eigenvalue decomposition of the target and interference SCMs:
\begin{equation}
    \mathbf{F}_{\mathrm{GEV}}(f) = \mathcal{P}\{\bm{\Phi}^{-1}_{\mathbf{NN}}(f)\bm{\Phi}_{\mathbf{XX}}(f)\}.
\end{equation}

While the target and noise masks estimated with the filterbank have a poor frequency resolution, this remains acceptable for estimating the target and noise SCMs, as the eigenvalue decomposition leads to the same eigenvector as long as the target SCM captures more energy from the target signal than the noise signal, and vice-versa.

Finally, as demonstrated by Heymann et al. \cite{heymann2015blstm}, a Blind Analytic Normalization (BAN) gain, defined as:
\begin{equation}
    g_{\mathrm{BAN}}(f) = \frac{\sqrt{\mathbf{F}^{\mathrm{H}}_{\mathrm{GEV}}(f)\bm{\Phi}_{\mathbf{NN}}(f)\bm{\Phi}_{\mathbf{NN}}(f)\mathbf{F}_{\mathrm{GEV}}(f)}}{D^2\mathbf{F}^{\mathrm{H}}_{\mathrm{GEV}}(f)\bm{\Phi}_{\mathbf{NN}}(f)\mathbf{F}_{\mathrm{GEV}}(f)},
\end{equation}
can be applied as a post-filter in order to reduce the distortion that can be introduced by the GEV beamformer in the target direction.
The enhanced STFT representation of the target signal is then obtained as follows:
\begin{equation}
    Z(t,f) = g_{\mathrm{BAN}}(f)\mathbf{F}^H_{\mathrm{GEV}}(f)\mathbf{Y}(t,f).
\end{equation}

\section{Results}

The algorithm's performance was evaluated by calculating the Signal-to-Distortion Ratio (SDR) improvement of the enhanced signal on a dataset containing 20 target speech tracks, 20 different speech tracks to use as interference, 20 ambient noise tracks with both stationary and non-stationary noise and 20 music tracks; of which every combination of target and interference was simulated on 5 different room impulse responses (RIRs). Each simulation was performed with a new RIR, with none being reused between configurations, for a total of 6000 unique RIRs.
The RIRs were simulated using the image method \cite{habets2006room} with the geometry of a ReSpeaker Core v2 microphone array as the receiver, as well as varying parameters as in the BIRD dataset \cite{grondin2020bird}, such as the room dimensions (width between $5$ and $15$ m, length between $5$ and $15$ m, and height between $3$ and $4$ m), absorption coefficients (between $0.2$ and $0.8$), speed of sound (between $340$ and $355$ m/sec), and target and interference source positions.
The speech segments are from the Librispeech \cite{panayotov2015librispeech} dataset, while the noise and music tracks were selected from the Musan \cite{snyder2015musan} dataset.

A value of $\gamma = 100$ is defined to separate between voiced and unvoiced spectrum segments (which represents a separation at $3125$ Hz), and $\alpha = 25$ is used as an adequate compromise between good target/interference separation and having enough data to properly calculate the SCM.

The mean SDR was calculated for the four following outputs: unprocessed mixture (from channel 1), delay-and-sum beamforming, KISS-GEV beamforming, and GEV beamforming with oracle ideal ratio mask (IRM) \cite{narayanan_ideal_2013}. The results of this experiment are shown in table \ref{tab:sdr_results}.

\begin{table}[!ht]
    \def\arraystretch{1.3}
    \centering
    \caption{SDR (in dB) in various noise conditions}
    \begin{tabular}{cccc}
    \hline\hline
        \multirow{2}{*}{Method} & \multicolumn{3}{c}{Types of interference}\\
        & Ambiant & Music & Speech \\
    \hline
        Unprocessed & 4.05 & 3.92 & 8.76 \\
        Delay-and-sum & 5.69 & 4.36 & 9.06 \\
        KISS-GEV & 10.31 & 10.42 & 13.09 \\
        GEV with oracle mask & 16.98 & 17.40 & 18.58 \\
    \hline\hline
    \end{tabular}
    \label{tab:sdr_results}
\end{table}

KISS-GEV significantly outperforms DS thanks to its ability to null interference. The remaining gap between KISS and Oracle masks is likely explained by the loss of spectral information to interference.

\begin{figure}[h]
    \centering
     \begin{subfigure}[b]{\linewidth}
         \centering
         \includegraphics[width=\linewidth]{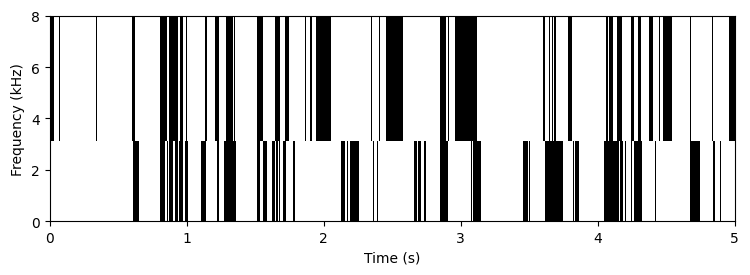}
         \caption{Binary mask generated with KISS-GEV}
         \vspace{10pt}
     \end{subfigure}
     \begin{subfigure}[b]{\linewidth}
         \centering
         \includegraphics[width=\linewidth]{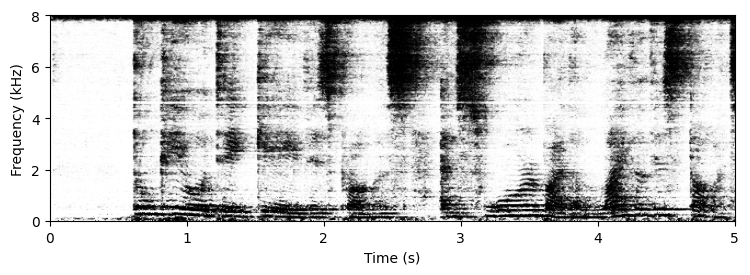}
         \caption{Oracle ideal ratio mask (IRM)}
     \end{subfigure}
     \caption{Time-frequency masks (the color black stands for a value of $1$, and white for a value of $0$)}
     \label{fig:masks}
\end{figure}

Figure \ref{fig:masks} contrasts the time-frequency mask generated with KISS-GEV with the Oracle IRM, on a single simulation with stationary noise. Apart from showing how a coarse mask still leads to effective results for SCM estimation, it also suggests dereverberation can be achieved, as it only takes into account the direct path from the target and ignores the late reverberation.

\begin{figure}
    \centering
     \begin{subfigure}[b]{\linewidth}
         \centering
         \includegraphics[width=\linewidth,height=0.3\linewidth]{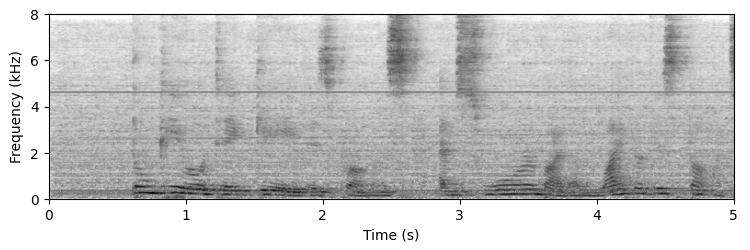}
         \caption{Unprocessed signal from channel 1 (SDR 11.65 dB)}
         \vspace{10pt}
     \end{subfigure}
     \begin{subfigure}[b]{\linewidth}
         \centering
         \includegraphics[width=\linewidth,height=0.3\linewidth]{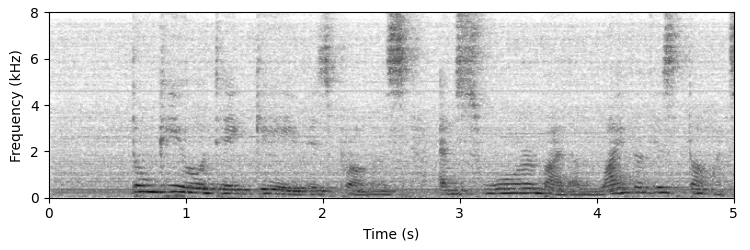}
         \caption{Delay-and-sum beamforming (SDR 13.54 dB)}
         \vspace{10pt}
     \end{subfigure}
     \begin{subfigure}[b]{\linewidth}
         \centering
         \includegraphics[width=\linewidth,height=0.3\linewidth]{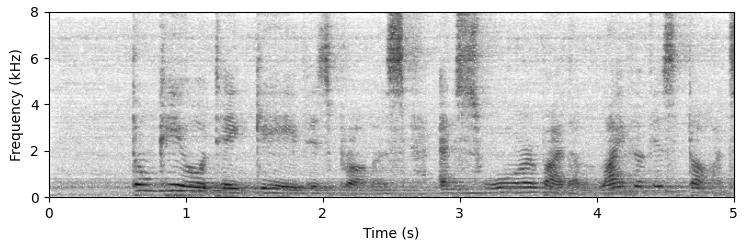}
         \caption{KISS-GEV beamforming (SDR 19.33 dB)}
         \vspace{10pt}
     \end{subfigure}
     \begin{subfigure}[b]{\linewidth}
         \centering
         \includegraphics[width=\linewidth,height=0.3\linewidth]{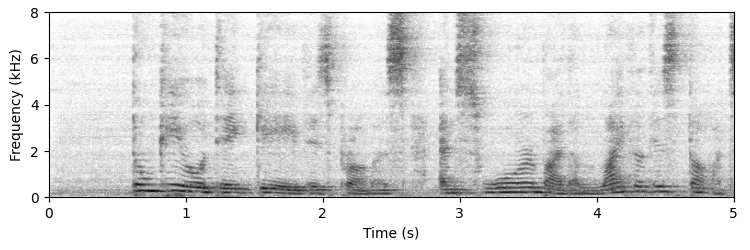}
         \caption{GEV beamforming with oracle mask (SDR 24.28 dB)}
     \end{subfigure}
     \caption{Spectrograms of the enhanced mixture, noise interference. Note the variation in the sharpness of upper harmonics (1.0 - 2.5 kHz) around 1 s mark and of fricative (4.0 - 8.0 kHz) at 2.5 s.}
     \label{fig:spectrograms_noise}
\end{figure}

Figure \ref{fig:spectrograms_noise} shows example spectrograms of the four outputs on the same simulation as figure \ref{fig:masks}. It can be observed that although it gives a similar noise floor to DS, KISS-GEV yields a much clearer resolution of the harmonics in voiced segments, as well as a more robust dereverberation than the other methods, which is especially visible on the fricatives. It also removes more white noise in the lower frequencies than DS.

\begin{figure}[t]
    \centering
     \begin{subfigure}[b]{\linewidth}
         \centering
         \includegraphics[width=\linewidth,height=0.3\linewidth]{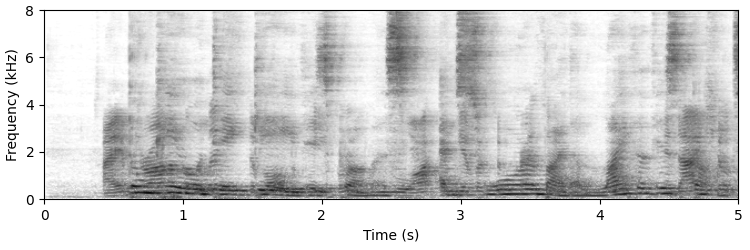}
         \caption{Unprocessed signal from channel 1 (SDR 7.86 dB)}
         \vspace{10pt}
     \end{subfigure}
     \begin{subfigure}[b]{\linewidth}
         \centering
         \includegraphics[width=\linewidth,height=0.3\linewidth]{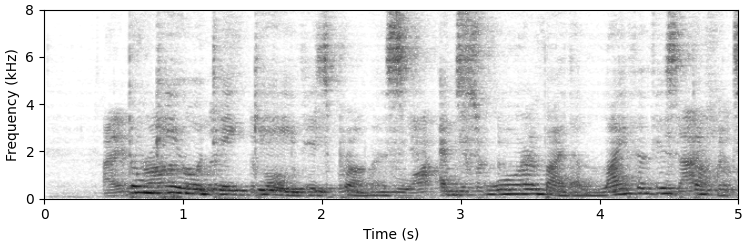}
         \caption{Delay-and-sum beamforming (SDR 9.77 dB)}
         \vspace{10pt}
     \end{subfigure}
     \begin{subfigure}[b]{\linewidth}
         \centering
         \includegraphics[width=\linewidth,height=0.3\linewidth]{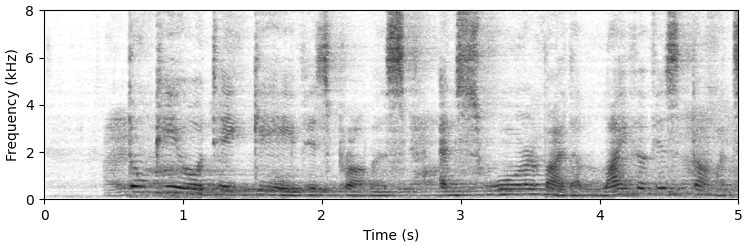}
         \caption{KISS-GEV beamforming (SDR 14.94 dB)}
         \vspace{10pt}
     \end{subfigure}
     \begin{subfigure}[b]{\linewidth}
         \centering
         \includegraphics[width=\linewidth,height=0.3\linewidth]{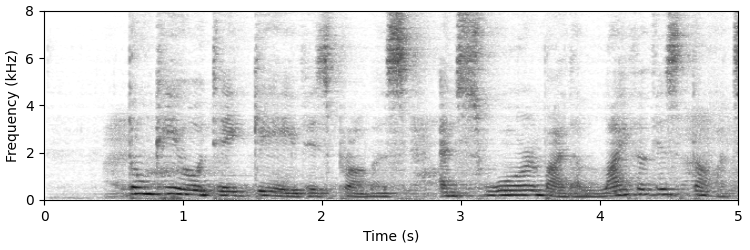}
         \caption{GEV beamforming with oracle mask (SDR 21.18 dB)}
     \end{subfigure}
     \caption{Spectrograms of the enhanced mixture, speech interference. Note the attenuation of voiced interference (0.0 - 4.0 kHz) segment before target utterance starts, around 0.6 s, and at 4.8 sec.}
     \label{fig:spectrograms_speech}
\end{figure}

Figure \ref{fig:spectrograms_speech} shows spectrograms of a simulation using speech as interference on the same target utterance as figure \ref{fig:spectrograms_noise}, where a sharp dereverberation is also demonstrated, while offering a significantly better attenuation of the interference voiced segments than DS.
This demonstrates that using KISS-GEV with informed DoA can enhance speech in the direction of interest and deal with the permutation issue observed in speech separation based on neural networks.

\section{Conclusion}

This paper presents KISS-GEV, a lightweight mask estimation front-end to generate target and interference SCMs for GEV-BAN beamforming that requires no training, hence performs well when enhancing against unseen interference.
Results show a significantly better SDR than the popular DS beamformer while relying only on the same target DoA assumption.
This approach is thus ideal for low-cost embedded hardware deployed in real-life environments.
As the end use of this method is speech processing, Word Error Rate (WER) improvement should also eventually be evaluated with ASR backend, such as Kaldi \cite{Povey_ASRU2011}.
Future work could also involve estimating the masks and SCMs in an online manner, for real-time applications.


\clearpage

\bibliographystyle{IEEEbib}
\bibliography{refs}

\end{document}